\documentclass{hcj}

\usepackage{natbib}

\usepackage{amssymb}
\usepackage{amsmath}
\usepackage{graphicx}

\usepackage{hyperref}
\usepackage{vaucanson-g}

\journalyear{2014}
\journalvolume{1}
\journalissue{1}
\journalpages{5-29}
\articledoi{10.15346/hc.v1i1.2}
\journalcopyright{\copyright{} 2014, DeDeo. CC-BY-3.0}
\title{Group Minds and the Case of Wikipedia}
\author{Simon DeDeo\affil{School of Informatics and Computing, Indiana University, Bloomington, IN 47408, USA \& Santa Fe Institute, Santa Fe, NM 87501, USA}}
\authorrunning{DeDeo}

\begin{document}

\maketitle

\begin{abstract}
Group-level cognitive states are widely observed in human social systems, but their discussion is often ruled out \emph{a priori} in quantitative approaches. In this paper, we show how reference to the irreducible mental states and psychological dynamics of a group is necessary to make sense of large scale social phenomena. We introduce the problem of mental boundaries by reference to a classic problem in the evolution of cooperation. We then provide an explicit quantitative example drawn from ongoing work on cooperation and conflict among Wikipedia editors, showing how some, but not all, effects of individual experience persist in the aggregate. We show the limitations of methodological individualism, and the substantial benefits that come from being able to refer to collective intentions, and attributions of cognitive states of the form ``what the group believes'' and ``what the group values.''
\end{abstract}

\section{Introduction}

Accounts of the history and structure of human behavior naturally attribute cognitive properties not just to individuals, but to groups. They do so despite strong methodological rules against such talk \cite{arrow1994}: fields as varied as political science, history, anthropology and economics make constant reference to the attitudes, expectations, beliefs, and values of groups. Scholars talk about the desires and strategies of a social class \cite{marxcurrents}, the expectations of financial markets \cite{kirman2010complex}, or the attitudes of a nation \cite{elias2000civilizing}. It is generally---usually implicitly---understood that talk about group-level mental states is not simple shorthand for talk about the mental states of individuals. The relationship between (for example) ``society's values'' and the behaviors and beliefs of its citizens, or ``the beliefs of the committee'' and those of its members, is still at best only partially understood.

Reference to group-level cognitive properties in the social sciences has thus been contentious \cite{wegner1987transactive}. Yet the idea of minds overlapping and nesting comes naturally to those in artificial intelligence, where mind is understood solely as computation. As far back as 1967, Marvin Minsky wrote that ``the question of where a particular machine ends and its environment begins can be settled only by a convention or definition'' (\citet{minsky1976computation}, Pg.~19) and urged that methodological criteria alone settle the question of where boundaries should be drawn.

The very success of the traditional psychological sciences points to the empirical utility of the biological boundary. In this article we provide a scientific account of how that boundary can be violated in informal, self-organizing social systems.

We shall show how reference to the irreducible mental states, and psychological dynamics, of a group mind is necessary to build sensible mathematical theories of large scale social phenomena. To do so, we provide an explicit example drawn from recent work on cooperation and conflict among Wikipedia editors \cite{dedeo2013collective}.

Section~\ref{jm}, ``Joint Machines and Social Variables'', introduces our account of group minds by reference to finite state machines. We show how the desire to describe coarse-grained properties of interacting machines leads naturally to the use of computational states that are disjunctive unions of the underlying mental states of the individuals involved.

Section~\ref{gm}, ``Group Minds in Wikipedia,'' provides an empirical example of how the considerations suggested by the toy models of Section~\ref{jm} play out the empirical dynamics of cooperation and conflict on Wikipedia. We build intuition for the problem by reference to a finite-state model of behavior among a large group of heterogenous agents, and find the Nash Equilibrium explicitly. We then show how reference to the properties of group minds is essential for a parsimonious account of real-world behavior. 

Section~\ref{beyond}, ``Subpopulations and Subminds,'' shows how, in Wikipedia, features of an individual's experience may, or may not, alter group-level accounts. Wikipedia users can interact in a variety of ways, including via discussion on article ``talk'' pages. Some forms of interaction leave strong traces at higher levels of organization, while others---including talk page interaction---coarse-grain away.

Section~\ref{conc}, ``Conclusions'', places these results in a larger context, connecting our work to recent work in neuroscience, complex systems, and political science. 

In two Appendices, we provide additional detail on our statistical methods, and on an interesting subtlety in the use of reverts to study conflict on Wiki-like systems.

\section{Joint Machines and Social Variables}
\label{jm}

Finite-state machines provide our first example of the transition from individual to group-level accounts of cognition.

Such machines model individuals as cognitively-limited transducers: conditional upon an environmental input and the current internal state, the individual produces a behavioral output and changes state. As the name suggests, the individual has only a finite number of possible internal states, and thus an explicitly bounded number of response-patterns. Such machines provide a highly simplified account of cognition, with internal states associated with a mental state or ``disposition''.

\begin{figure}
\includegraphics[width=3.5in]{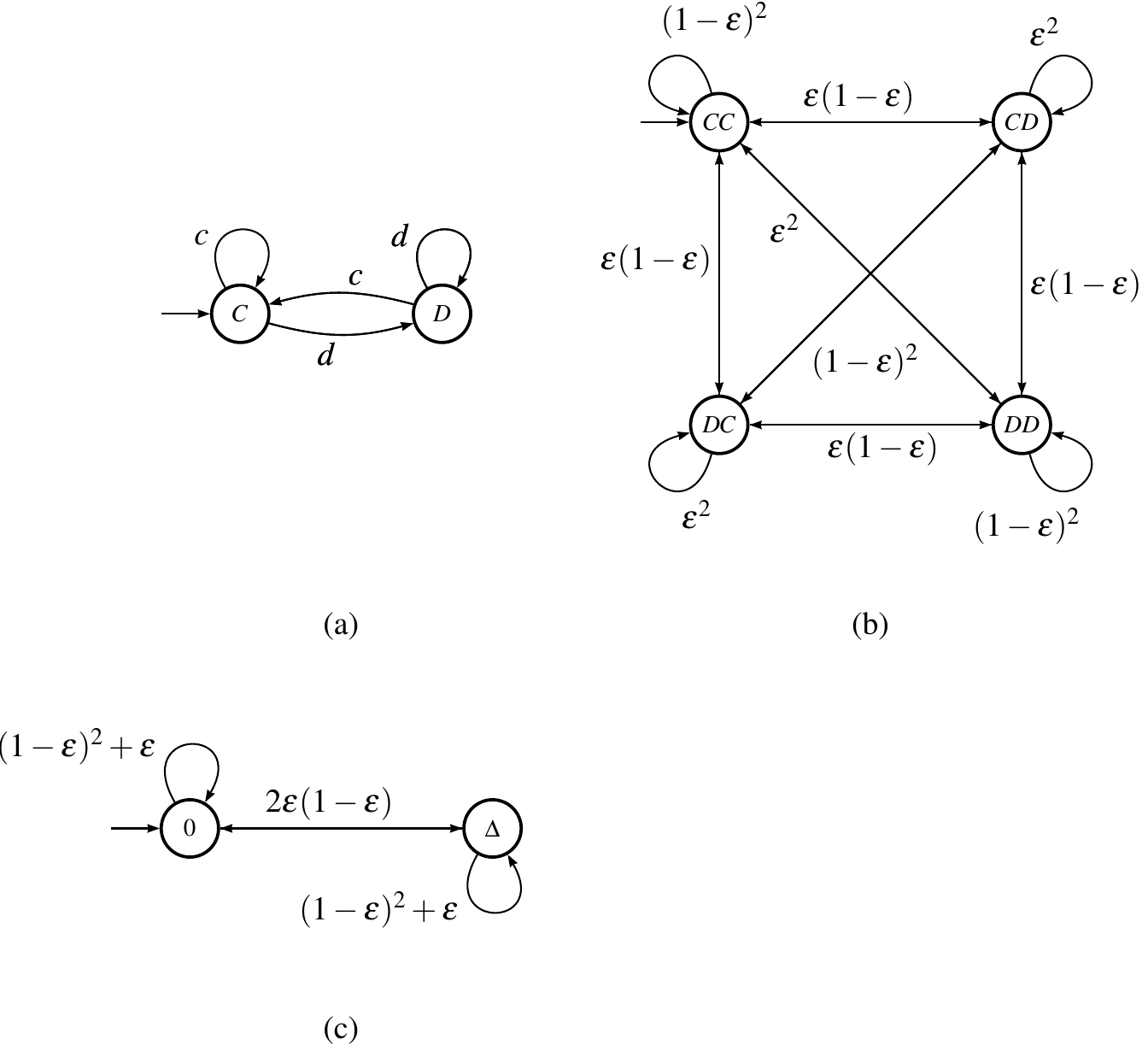}
%
%
%
%
%
%
%
%
%
%
%
%
%
%
%
%
%
%
%
%
%
%
%
\caption{{\bf Machines, joint machines and effective theories.} {\rm (a): a finite-state machine representing the ``tit for tat'' (T4T) strategy in prisoner's dilemma. The node labels the move the agent makes; the arrows label the transitions between states made by the agent upon observing the behavior of its opponent. (b): the joint machine describing the interaction of two T4T agents playing in a noisy environment; nodes label the observed behaviors, while arrows are labelled with transition probabilities. We assume a binary symmetry channel with noise, such that a {\tt c} (respectively, {\tt d}) is flipped to a {\tt d} (respectively, {\tt c}) with probability $\epsilon$. (c): a coarse-grained machine describing the evolution of a social variable tracking inequality of outcome; state 0 when both agents receive the same payoff, and state $\Delta$ when one receives a higher payoff than the other due to unreciprocated cooperation.}}
\label{t4t}
\end{figure}
The famous ``tit for tat'' (T4T) solution for iterated prisoner's dilemma is a particularly simple example. Written as a finite-state machine, T4T has two internal states, {\tt C} and {\tt D}, which both name the internal state and dictate the machine's next action, cooperate or defect. Depending on the action previously observed, it shifts between these two states (Fig.~\ref{t4t}a). An opponent's defection ({\tt d} signal) in the current round shifts the machine to state {\tt D}, and it will defect in the next round.

\subsection{Joint Machines and Computational Macrostates}

The T4T solution gained early attention in the quantitative study of altruism and social cooperation. It did so both because of its natural interpretation in terms of (stylized) facts about social conventions, and because of its behavior under noise: T4T is ``forgiving'', and allows opposing strategies to recover from an accidental defect.

This can be seen in Fig.~\ref{t4t}(b), where we show the \emph{joint machine} that emerges when two T4T strategies interact on a noisy channel. Each state of Fig.~\ref{t4t}(b) corresponds to a joint specification of the underlying computational states of the two individuals. With some probability $\epsilon$, a transmitted symbol is flipped to its opposite; it is this flipped symbol that is both received by the other machine and determines the payoff in that round. Noise can send a cooperative pair into a cycle of alternating, or even mutual, defection, but with non-zero probability of recovering the cooperative solution.

In the T4T joint machine, the states of the system are also causal states of the fine-grained system \cite{crutchfield1989inferring}; knowledge of the current play ($CC$ vs.~$CD$, and so forth) tells us everything we need to know about the future probabilities of action.

When it comes to building accounts of social systems, we are necessarily concerned with aggregate properties; a theory of social systems is not an account of the complete psychological states of the individuals, but is rather phrased in terms of social variables that summarize the underlying states. As an example, we may be interested not in the dynamics of the individual psychological states of the T4T machines, but only in the (absolute) inequality of outcome. In prisoner's dilemma, the absolute difference in payoff is either zero (for $CC$ and $DD$---both ``win'' or both lose) or $\Delta$ (greater than zero; in the case where one player receives the payoff $w>0$, and the other player receives the ``sucker's payoff'', $s<0$).

These coarse-grained equivalence classes define \emph{macrostates} \cite{shalizi2003macrostate} of a higher-level theory; the dynamics of these macrostates are effective theories \cite{dedeo2011effective}. In the joint-T4T case, the macrostates are, again, causal states: knowledge of the current payoff inequality tells the observer everything she needs to know about the future behavior of the system. (Macrostates need not be causal states, as can be seen in for a $0$/$\Delta$ coarse-graining in the interaction of two ``tit-for-two-tats'' (T42T) strategies, where T42T is defined as ``defect if and only if defected against twice in a row.'')

A coarse-graining that refers unambiguously to the state of one, and only one, of the two individuals---\emph{e.g.}, a coarse-graining to $CC$-or-$CD$ and $DC$-or-$DD$---does \emph{not} provide causal states and thus can, at best, provide only imperfect predictions of future behavior. By contrast, it is easy to construct situations in which the $0$/$\Delta$ macrostates---and only these macrostates---are relevant to a third-party observer. For example, one can augment the prisoner's dilemma with a third player who ``wins'' if she correctly picks the inequality level. In this case, the third player never has cause to refer to states of the individual players; she need only refer to disjunctive unions ($CC$-or-$DD$ vs.~$CD$-or-$DC$).

The desire to talk about coarse-grained properties of a system leads us to theories whose basic units are disjunctive unions of joint mental states. In this simple case, there is no need to refer to the properties of individuals; not only is our theory simpler when we work at the group-level, we lose no predictive power.

\subsection{From Machine to Mind}

The previous section provided a toy example of the Minsky intuition: that computational accounts of behavior may remain agnostic about the boundaries between interacting systems. We have further shown how the social variables relevant to a third agent may mean that the relative computational states of the system do not uniquely and separately identify the mental states of individuals.

Finite-state accounts, however, leave much to be desired. They provide no real space for intentionality, or the ``aboutness'' of internal states; it is only by strained analogy, for example, that one can claim a finite-state machine believes something about its environment.

In addition, our use of only two (or three) individuals allows us to demonstrate only a very limited gain from reference to joint mental states. Let us take an informal example. One might refer to a store as ``friendly'' if at least one of its two clerks decides to help. But such an attribution, of a mental state (willingness to help) to a group agent (``the store''), provides little gain over and above the statement that ``at least one of the two clerks was willing to help'', and this latter statement makes no reference to mental properties other than those of the individual agents. 

Finally, it is hard to see in these simple examples how someone could come to know the macrostate without coming to know the microstate. The reason I call a store friendly is precisely because I've observed a ``normal'' mental property of one of the clerks, and similar objections would appear to apply in any natural account of coming to know the macrostates of Fig.~\ref{t4t}(c).

When both the number of individuals and the space of possible behaviors is small, in other words, talk of group minds may be easily schematized---but provides only limited benefit. In the next section, we will see how they become essential in the description of large-scale social phenomena.

\section{Group Minds in Wikipedia}
\label{gm}

In this section, we will demonstrate that observed behavior on Wikipedia urges us to postulate the existence of group-level cognitive states and laws.

We first present a game-theoretic analysis of conflict on Wikipedia. We then show how the observed phenomenology requires us to modify and extend the simple assumptions of this game. We show how these extensions lead us to an account of group cognition that is not simply reducible to that of the individuals involved.

\subsection{The Wikipedia System}

Founded in 2001, the online encyclop\ae dia Wikipedia has been a source of scholarly attention for over a decade \cite{bar2014twelve}. While the legally-recognized Wikimedia Foundation has on the order of 100 employees, day-to-day behavior of the $10^5$ volunteer editors is guided by a set of overlapping policies, guidelines, permanent and \emph{ad hoc} committees, formal and informal rules, a range of behavioral norms and cultural practices, and (perhaps most importantly) the pragmatic expectations pseudonymous users form of each other.

Such a large population ends up a cohesive, though far from conflict-free, culture that has been the source of ethnographic studies \cite{reagle2010good,jemielniak2014common}, interventional experiments (\emph{e.g.}, \citet{regulating, kittur}) and quantitative observational studies (\emph{e.g.}, \citet{tails,roles,greenstein2012wikipedia}). From around 2006 to 2012, the number of ``active'' volunteer editors ranged between 20,000 and 60,000; the plateau and then decline in the number of users identified by \citet{halfaker2012rise} points to the long-term evolution of a project that six years previously underwent rapid growth.

\subsection{Social Norms, Page Reverts, and a Square-Root Law}

To study the characteristics of this system, we focus on the page-by-page time-series of ``reverts'' in the encyclopedia. Reverts---informally, when one editor ``undoes'' the work of another---are an excellent tracer of short-term conflict \cite{kittur2010beyond}; they also play a central role in the social norms of the larger Wikipedia community, where users discuss ``Edit Warring'' (repeated undoing of other's work) in depth, and the ``three revert rule'' (do not revert more than three times on a single page) makes direct reference to the behavior as a bright line for policy violation.\footnote{See \url{https://en.wikipedia.org/wiki/Wikipedia:Edit_warring} (last accessed 3 July 2014). For highly-edited pages, the main exception to non-reverting as a social norm is the case of ``obvious vandalism'': editors are encouraged to revert contributions that vandalize a page, where vandalism is defined as a ``deliberate attempt to compromise the integrity of Wikipedia. Examples of typical vandalism are adding irrelevant obscenities and crude humor to a page, illegitimately blanking pages, and inserting obvious nonsense into a page''. See \url{https://en.wikipedia.org/wiki/Wikipedia:Vandalism}, last accessed 6 July 2014. How the community handles vandalism provides a source of background noise to our study---in particular, we are ``off by one'' in counting the position of the norm-violating action. Vandalism levels are influenced by ``page protection'', a top-down action that can restrict the editing of articles to different user classes. Early data on protection was not systematically logged, but from 23 December 2004 onwards, Wikipedia software maintained records via version 1.4 of the MediaWiki software (see \url{https://en.wikipedia.org/wiki/Wikipedia:Village_pump_(technical)/Archive_104}, last accessed 6 July 2014). The protection status of the {\tt George\_W.\_Bush} page changed 243 times between then and 3 July 2014; over the same period, the page saw 6,339 non-reverting runs. At most, therefore, less than 4\% of non-reverting runs had continuation probabilities influenced by changes in top-down regulation.} Though necessarily imperfect, automated identification of reverts allows us to track sites of conflict over $\sim10^6$ individual actions by $\sim10^5$ editors,\footnote{For the 979,426 editing actions in our set, we have 224,080 ``users''; despite Wikipedia norms that encourage the use of a single, named account, 140,196 are identified only by IP address, and 83,884 by username.} spread over a set of 62 highly-edited pages.

We focus on the statistics of ``continued non-reversion'' which we refer to informally as ``cooperative runs''. In particular, we study how the probability that the next edit will be a revert declines as a function of time since the last revert. The phenomenological form of this law was found in \citet{dedeo2013collective} to be
\begin{equation}
P(R|RC^k) = \frac{p}{(k+1)^\alpha},
\label{phenom}
\end{equation}
where $P(R|RC^k)$ is the probability that one sees a revert edit after exactly-$k$ non-reverting edits, and $p$ and $\alpha$ are constants. Empirically, $\alpha$ ranges between 0.4 and 0.8, with an average (in our current data-set) of $0.5\pm0.1$, an approximate inverse square-root scaling of propensity to revert with length of cooperative run.

We are concerned only with how that small section of history, from the recent revert to the current edit, affects behavior; explicitly, our measurements average out over variety of larger contexts that change on longer timescales. This allows us to build up sufficient statistical power to study the detailed structure of the conditional behavior described by Eq.~\ref{phenom}. 

\begin{figure}
\includegraphics[width=3.5in]{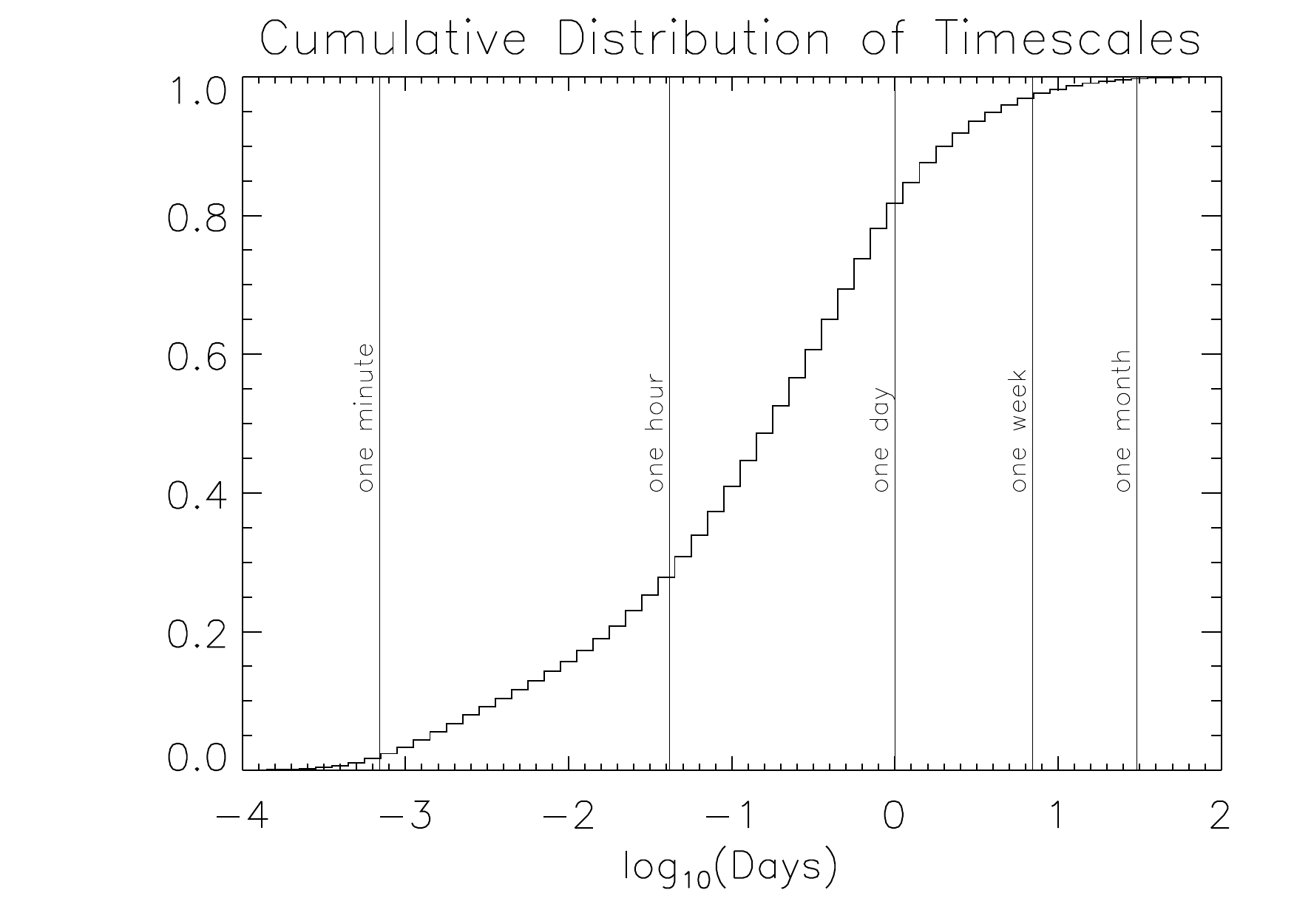}
\caption{{\bf Cooperation is rapid...} {\rm Most cooperative runs last less than twenty-four hours. Plotted is the cumulative distribution of clock-times, for $171,107$ $RC^kR$ runs measured on 62 highly-edited pages, showing that the timescales over which we measure behavioral changes are usually much shorter than day-to-week timescales of exogenous events, and the week-to-year timescales of Wikipedia policy changes.}}
\label{timescale}
\end{figure}
The timescales of these runs are indeed short. Fig.~\ref{timescale} shows the cumulative distribution of timescales of cooperation for a sample of $171,107$ cooperative runs measured over 62 highly-edited pages. Although occasional long-tail events are possible, more than 80\% of the time, a page in our sample can not go more than a day without a revert. Only a small number of cooperative runs last longer than a week; many of these observations come from early in the encyclopedia's history, and concern less controversial topics. On the other extreme, a small fraction of cooperative runs (2.5\%) are over (\emph{i.e.}, terminated by a revert) in less than a minute. 

\begin{figure}
\includegraphics[width=3.5in]{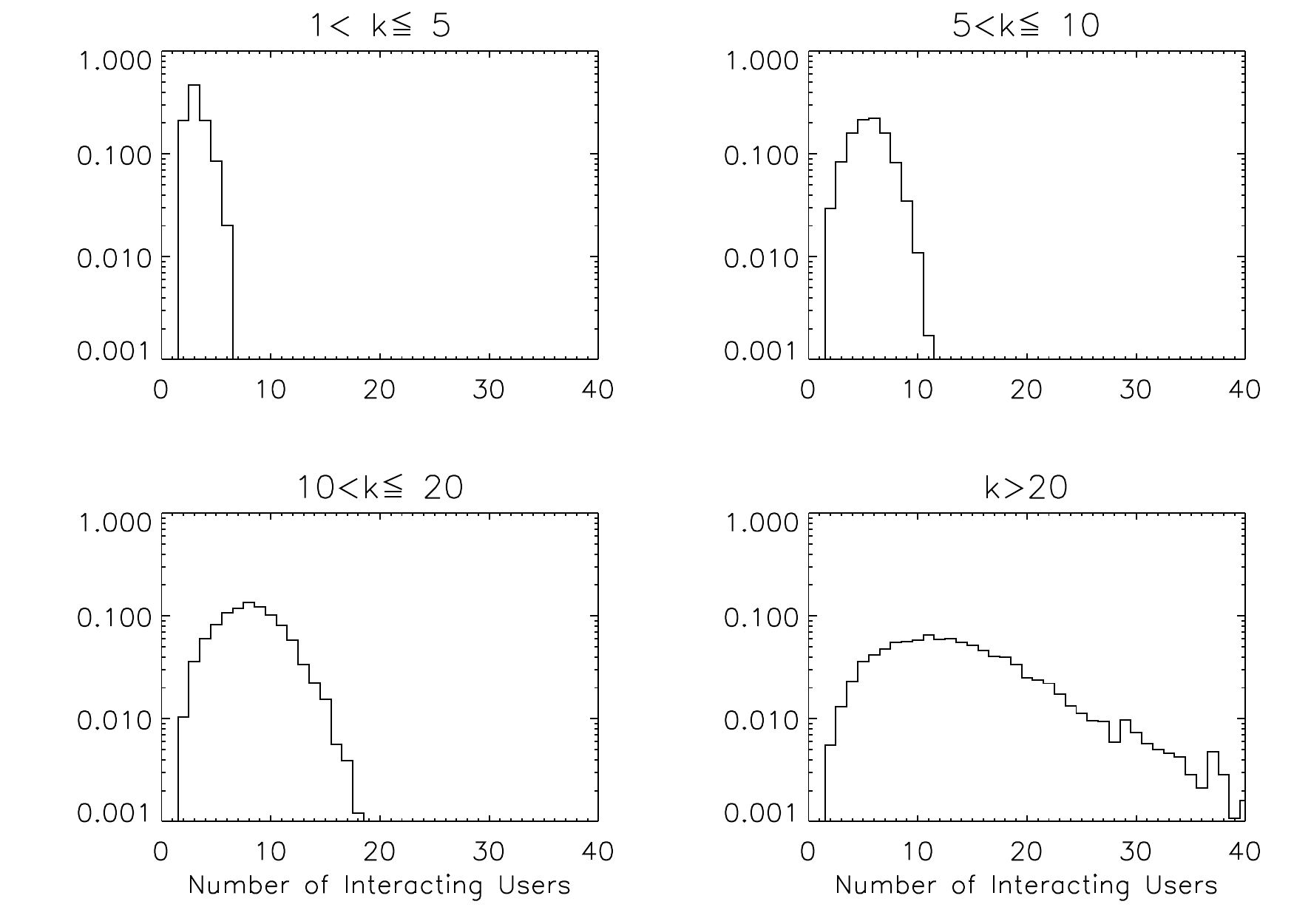}
\caption{{\bf ...and social}. {\rm Most cooperative runs involve multi-user interactions. Over 85\% of all runs with $k>1$ involve at least three parties; for longer runs, $RC^kR$ with $k$ greater than five, over 97\% of interactions involve at least three parties. The very longest runs, $k>20$, involve a median of twelve editors. Here we count the unique users responsible for the $k$ $C$ edits and the terminal $R$, so that (for example) a run of length $k=2$ can have at most three unique editors.}}
\label{num_users}
\end{figure}
Cooperative events are not only reasonably rapid, but also highly social. Fig.~\ref{num_users} shows the distribution of the number of users involved in these runs. For $k>1$, over 85\% of all runs involve at least three users, and the longest runs can draw in dozens. Interactions over and above the commonly-studied dyadic, or pair-wise, case are essential to the phenomenology we observe. 

\begin{figure}
\includegraphics[width=3.5in]{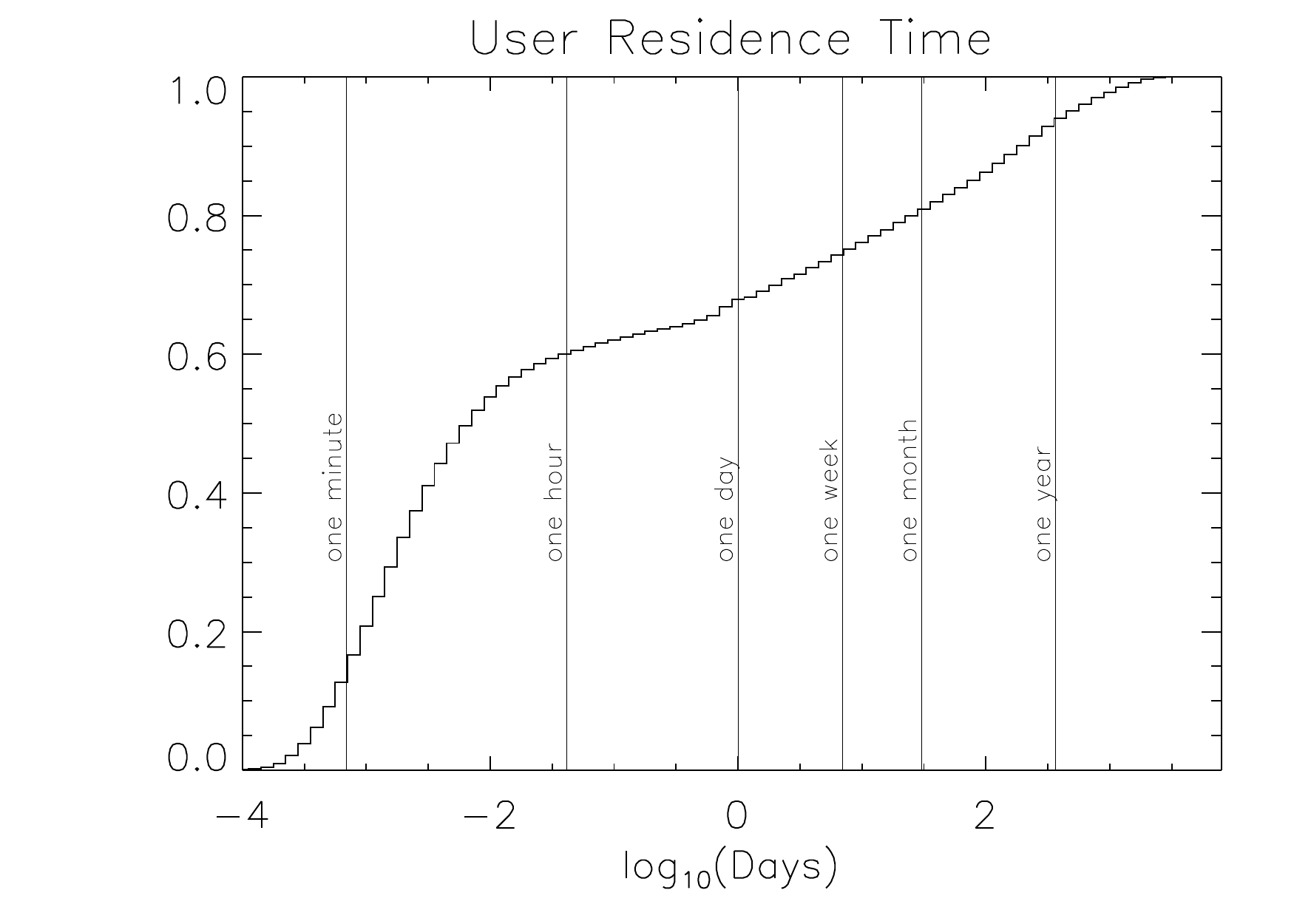}
\caption{{\bf The majority of users have only short-term engagement with a page.} {\rm Plotted here is the cumulative distribution of user longevity---the length of time that elapses between the first and last observed edit on a page---over 62 highly-edited pages. More than half of users have a total engagement time of less than a day; only a small fraction (less than 10\%) of users are engaged with editing a page for more than a year.}}
\label{long}
\end{figure}
This sociality involves users who may engage for only short periods. Fig.~\ref{long} shows the cumulative distribution of ``residence times''---the elapsed time between the first and last edit made by a particular user on a particular page. While the timescale of user engagement is broad, spanning nearly eight orders of magnitude, more than half of users have a total engagement time of less than a day.

We will return to these three facts---concerning the rapidity, sociality, and turnover of the community---when we come to modify and extend the simple game-theoretic account we present in the following section.

\subsection{The Assistance Stage Game for Wikipedia}
\label{asg}

This section presents a game theoretic model of revert behavior on Wikipedia. Such a rational-choice account will provide a guide for how to map, in a rigorous fashion, cognitive states---beliefs about the world---to actions. As we shall see, these standard models fall short of describing real-world behavior: they require us to believe implausible facts about individuals.

We model the decision problem faced by users as a game of perfect, but incomplete, information \cite{kreps1990game}. Our model is emphatically \emph{minimal}: rather than propose a full account of the motives and contexts for individual editors, we construct a parsimonious account consistent with the most basic facts about interaction on Wikipedia.

In particular, we introduce three important facts that constitute distinctive features of Wikipedia-like systems. 

{\bf Firstly}: no top-down agency chooses which user next edits a page. The order in which users edit, and the identity of who will edit next, is uncertain. In this first analysis we approximate the arrival of users as an {\sc iid} process.

{\bf Secondly}: the page history is visible to users; individuals are aware when those who come before have chosen to revert. This defines the information available to users.

{\bf Thirdly}: individuals have only incomplete information about the motives and desires of their fellow users. We model this incompleteness as an uncertainty about whether the next user will perceive the revert/don't-revert choice as (on the one hand) a problem of misaligned incentives or (on the other) a problem of aligned incentives and mutualism.

While these facts capture the sources of knowledge and uncertainty most relevant to system function, they are only imperfectly observed in the real world. In Sec.~\ref{beyond}, below, we critically evaluate the relaxation of the third, and strongest, assumption.

For now, given these assumptions, we construct an extensive-form stage game with a randomly-selected individual making a single choice: to revert (R), or cooperate (C). Her payoff for this move is the sum of two games, one played with the individual who edited just-previously, and the other played with the next editor to appear. The payoff matrix of each game is of the form
\begin{equation}
\begin{tabular}{c|cc}
 & C & R  \\ \hline
C & $c+\epsilon\theta_1$, $c+\epsilon\theta_2$ & $s+\epsilon\theta_1$, $w$ \\
R & $w$, $s+\epsilon\theta_2$ & $0$, $0$
\end{tabular}
\end{equation}
where $\epsilon$ is a constant, $0<c<w$, and $s<0$. We model a heterogeneous population of players using $\theta_1$ and $\theta_2$, independent random variables drawn uniformly between zero and unity. These are chosen once (and for all time) for each player, and boost the payoff associated with the normative action.

As an example, say that the player in question, Alice, chooses to not to revert; say further that the previous player, Walrus, also chose not to revert; say finally that the player after Alice, Carpenter, responds to Alice by choosing to revert. Alice's payoff is $(c+\epsilon\theta_1)+(s+\epsilon\theta_1)$. This is her $CC$ payoff from cooperating with Walrus, plus the negative ``sucker's payoff'', $s$, from being reverted by Carpenter. On making her move, Alice is aware of Walrus's choice, but does not know the identity (or move choice) of the next player.

Depending on the value of $\epsilon$ and $\theta_1$, Alice's payoff might be positive (because of her cooperation with Walrus, and a general desire to follow social norms against reverting), or negative. Meanwhile, we do not know what Carpenter's final payoff will be until the player after him makes a move, but we know that it is either $w$ (if the next person reverts) or $2w$ (if the next person cooperates).

The random variable governs the uncertainty a user has about her fellow-user's perception of the game; to be clear, Alice knows her $\theta$ value (and thus whether she likes to revert others, or prefers to find a cooperative solution), but does not know the character of the next editor. When formulating her strategies for play, therefore, she will have to estimate her utility in the face of uncertain knowledge of others.

\subsection{What Alice Believes about Walrus and the Carpenter}
\label{alice-belief}

We can make Alice as intelligent as we like, but what sets the complexity of her behavior are the beliefs she holds about the people she encounters.

We shall (at first) give Alice uncharitable views of the sophistication of her fellow editors. In particular, we shall have her believe that editors condition their behavior solely on whether the previous editor did, or did not, revert. She has, in other words, two beliefs: $\beta_c$ (the probability that the next editor will revert her, given that she cooperated) and $\beta_r$ (the probability that the next editor will revert her, given that she reverted).

Let us say the previous editor, Walrus, cooperated. Then Alice's best response (she will reason) is to cooperate if and only if
\begin{equation}
(c+\epsilon\theta_1)+[(c+\epsilon\theta_1)(1-\beta_c) + (s+\epsilon\theta_1)\beta_c)] > w + w(1-\beta_r).
\label{alice}
\end{equation}
This inequality should be easy to read; the first term (on the {\sc lhs}) is the benefit to cooperating with Walrus; the second term (in square brackets) is the expected benefit to cooperating with the next editor, given expectations about the next editor's strategies. The third and fourth terms (on the {\sc rhs}) are (respectively) the (now known) payoff from reverting Walrus, and the expected gain given the possibility that the next user will cooperate (with probability $1-\beta_r$).

Conversely, and by similar reasoning, if Walrus reverted, Alice's best response is cooperation if and only if
\begin{equation}
(s+\epsilon\theta_1)+[(c+\epsilon\theta_1)(1-\beta_c) + (s+\epsilon\theta_1)\beta_c)] >  w(1-\beta_r).
\label{alice2}
\end{equation}

If Alice has no way of predicting the identity of the next editor, and (furthermore) if she truly believes her fellow editors can hold no more sophisticated a memory than $C$-vs-$R$---among other things, for example, if she believes that editors do not respond to, or cannot perceive, ``punishment''---then she might as well choose the action that most benefits her. In other words, Alice herself will adopt the strategy ``cooperate after cooperation'' if and only if her $\theta_1$ is such that the inequality of Eq.~\ref{alice} is satisfied; similarly, she will cooperate with a reverter only if the inequality of Eq.~\ref{alice2} is satisfied.

The final stage of the argument is to find the fixed point where beliefs about the system lead to actions that maintain the truth of those beliefs.\footnote{Game Theorists will recognize this as a Markov Perfect Equilibrium \cite{fudenberg1991game}.} If we ask the population as a whole to reason as Alice does, then the probability that the inequality Eq.~\ref{alice} holds is just the probability that $\theta$ for the individual in question is large enough, in particular that
\begin{equation}
P\left(\theta_1 > \frac{2(w-c)+\beta_c(c-s)-\beta_rw}{2\epsilon} \right),
\end{equation}
and for beliefs to be consistent with the actions observed, this probability has to equal $1-\beta_c$. A similar argument applies to the satisfaction of Eq.~\ref{alice2}, and, with two equations and two unknowns, we can solve for the self-consistent solution to the beliefs $\beta_c$ and $\beta_r$. The general solution is the ratio of polynomials; for the particular case where the disgruntlement, $s$, on being reverted is equal to $c-w$, $w-c$ is greater than zero and less than $\epsilon$, and $\beta_c$ and $\beta_r$ are equal, the expected behavior is particularly simple,
\begin{equation}
P(R|C)=P(R|R)=P(R|RC^k)=\frac{w-c}{\epsilon},
\label{markov}
\end{equation}
where $(w-c)/\epsilon$ is a constant.

We emphasize that Eq.~\ref{markov} is \emph{clearly incompatible} with the observed phenomenology, Eq.~\ref{phenom}. This is to be expected: the assistance stage game makes assumptions that we believe are naturally violated in the real world. 

However, Eq.~\ref{markov} does give us a clear account of how we can alter the model by hand. All we have to do is make $\epsilon$ a function of $k$---indeed, a very simple function of $k$, $\epsilon(k)\propto (k+1)^\alpha$---and we can recover Eq.~\ref{phenom} instantly. Such systematic shifts in $\epsilon$ can not be accounted for by the stage game we present above; to explain them we must violate some or all of the assumptions above.

In the next section, we discuss exactly what this violation means for our understanding of cooperation on Wikipedia, and the ways in which we describe the nature of belief and action in a large-scale social system.

\subsection{Three Theses for a Group Mind}
\label{complex}

Changing $\epsilon$ amounts to doing two things simultaneously. First, it amounts to changing the overall balance in the system: shifting agents away from seeing Wikipedia as a series of one-shot prisoner's dilemma games, and towards a mutualistic viewpoint that sees cooperation, even at risk of revert, as intrinsically good. Second, and equally importantly, it amounts to changing the beliefs that individuals have about others. This simple argument, based on long-understood features of game-theoretic models, suggests the first thesis of this section.

\begin{quote}
{\bf Thesis one}. For behavior to shift in the fashion we observe, both beliefs and desires must change in a systematic fashion.
\end{quote}

Belief change with $k$ will naturally occur in the real world. We expect that real Wikipedia users are playing sophisticated strategies, with potentially greater look-back times and sensitivity to features of the system such as the usernames, talk pages, and user pages of the editors currently active. Over the minutes and hours that elapse during the unbroken non-reverting sessions we examine here, they will naturally---and in violation of the simple assumptions of the stage game---come to change their future expectations.

We also expect preference change among Wikipedia users. Such change may be endogenously driven---I may simply enjoy cooperating more when others around me appear to be cooperating---or may be due to complicated mechanisms involving reputation effects that we know to be in effect in other public goods games.

Both beliefs and preferences will be driven by the changing state of the page itself. Pages will improve (or degenerate) over time, and the evolving state of the page will drive users (or their programmers, in the case of bots) towards (or away) from cooperation. The {\sc iid} assumption will fail as changing conditions shift the composition of the pool of active editors.

The rapid formation of these cooperative runs (Fig.~\ref{timescale}) suggests that these effects are unlikely to arise from a top-down process; the involvement of multiple users (Fig.~\ref{num_users}) allows for multiple distinct mechanisms for belief and preference change, and a variety of outcomes. The timescales of community turn-over for any particular page, shown in Fig.~\ref{long}, further suggests that users do not have sufficient time to model each other one-by-one. Both theoretical and empirical considerations, therefore, suggest we adopt a second thesis:

\begin{quote}
{\bf Thesis two}. The cognitive laws we can infer from Eq.~\ref{phenom} do not simply reduce to direct accounts of mental states of individual editors.
\end{quote}

Sec.~\ref{alice-belief} showed us how to infer beliefs and preferences from actions; but both mechanistic and theoretical considerations make it unlikely that these beliefs (``the next person will revert with probability proportional to $\epsilon(k)$'') and preferences (``my desire to follow social norms experiences an increase proportional to $\epsilon(k)$'') correspond to the beliefs or desires of any user whatsoever. 

In other words: Alice, Walrus and the Carpenter may be useful explanatory devices, but they do not correspond to any particular users in the real world. Real users are more limited (in the extent to which they can gather information about behavioral patterns), more nuanced in their desires, and likely attribute greater cognitive complexity to each other. However, knowledge and simulation of this reality is unnecessary to describe the behavior we observe.

Together, these two theses---a positive thesis connecting behavior to cognitive state, and a negative thesis preventing direct identification of those cognitive states with those of any individual---urge us to accept a third thesis, and the main result of our paper:

\begin{quote}
{\bf Thesis three}. The reliability of Eq.~\ref{phenom} should be explained by the existence of psychological laws pertaining to group-level mental states.
\end{quote}
Different disciplines have different traditions for talking about mental states and laws of this form. Economics, as mentioned above, has little difficulty referring to the beliefs of ``the market'' without necessarily attributing these beliefs to any participant \cite{hayek1945use}. Political scientists study core features of civil conflict by reference to group-level values and ideologies, even (and, indeed, particularly) when these conflict with the beliefs of group members \cite{Sanin01032014}. Other fields, such as intellectual history, may talk about such states by reference to an idealized individual, whose cognitive features are understood by comparison to biographical accounts of actual thinkers \cite{hazard2013crisis}.

\section{Subpopulations and Subminds}
\label{beyond}


Wikipedia norms urge that editor-editor communication take place on the associated talk page of the article in question;\footnote{See \url{http://en.wikipedia.org/wiki/Wikipedia:Talk_page_guidelines} (``The purpose of a Wikipedia talk page ... is to provide space for editors to discuss changes to its associated article or project page'') and \url{http://en.wikipedia.org/wiki/Wikipedia:Etiquette} (``When reverting other people's edits, give a rationale for the revert (on the article's talk page, if necessary), and be prepared to enter into an extended discussion over the edits in question''). Both pages last accessed 12 September 2014.} meanwhile, users can, and sometimes do, make edits to a page numerous times in a short period. Both of these practices provide information to others about an editor's goals, desires, and level of cooperativity over and above that possible in the stage game above.

To see if these micro-level facts ``bubble up'' to alter large-scale features of behavior, we can look at how behavior shifts within the subgroup of users who have recently interacted. To do this, we can look at the conditional behavior of this group, using the same phenomenological law of Eq.~\ref{phenom}. Taking ``recent'' interaction to mean ``within the last ten edits on the page,'' we can define
\begin{equation}
P(R|RC^k,\mathrm{int}) = \frac{p_\mathrm{int}}{(k+1)^{\alpha_\mathrm{int}}}
\label{phenomint}
\end{equation}
as the probability that, when a user who has edited the page within the last ten edits performs a \emph{new} edit, that edit is a revert. If the parameters that define $P(R|RC^k,\mathrm{int})$ and $P(R|RC^k)$ differ, this is a signal that the effective mental states of these subgroups evolve differently. 

\begin{figure}
\includegraphics[width=3.5in]{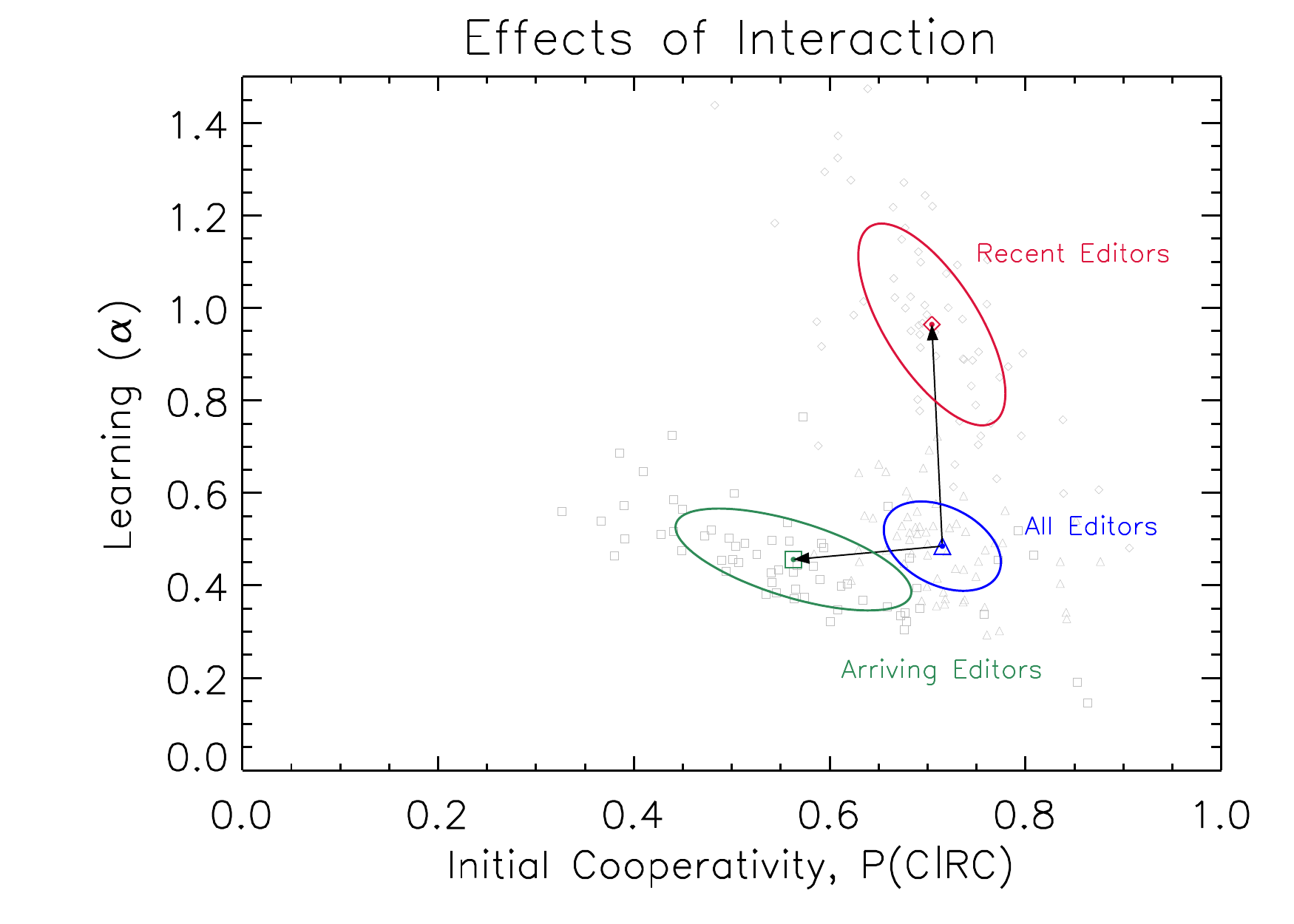}
\caption{{\bf Users who edit learn faster}. {\rm Compared to the population as a whole, users who have edited recently (within the last ten edits) begin with similar predispositions (initial cooperativity, $P(C|RC)$) but shift towards cooperative behavior more rapidly than others. Meanwhile, newly arriving editors tend to be initially less cooperative, and while they do (at the population level) learn to become more cooperative ($\alpha$ greater than zero), they do so more slowly than those who interact. Shown: distribution of $p$ and $\alpha$ parameters for 62 pages in our data, for all users (triangles; blue $1\sigma$ range), users who have recently edited (within the last ten edits; diamonds; red $1\sigma$ range) and newly-arriving users (squares; green $1\sigma$ range).}}
\label{int_edit}
\end{figure}
Fig.~\ref{int_edit} shows, for the 62 pages in our data set, how $\alpha$ and $P(C|RC)$ shift when one considers the population as a whole as compared to two sub-populations: editors who have recently (within the last ten edits) made an edit, and those who have not.

Parameters for the two sub-groups show a clear separation. When restricting to those editors who have recently interacted, one finds a similar $P(C|RC)$---\emph{i.e.}, a similar disposition at the beginning of a cooperative streak---but a much larger $\alpha$. Instead of the square-root law, we find instead that the probability of a revert declines roughly as $1/k$. Informally, group-level dispositions for editors who interact begin similarly, but shift more rapidly towards cooperation. We emphasize that this ``learning'' describes a process that occurs at the group level---any individual editor has a complex set of beliefs and desires, but their effects can be explained by reference to a collective mental state of the population of which she is a member.

\begin{figure}
\includegraphics[width=3.5in]{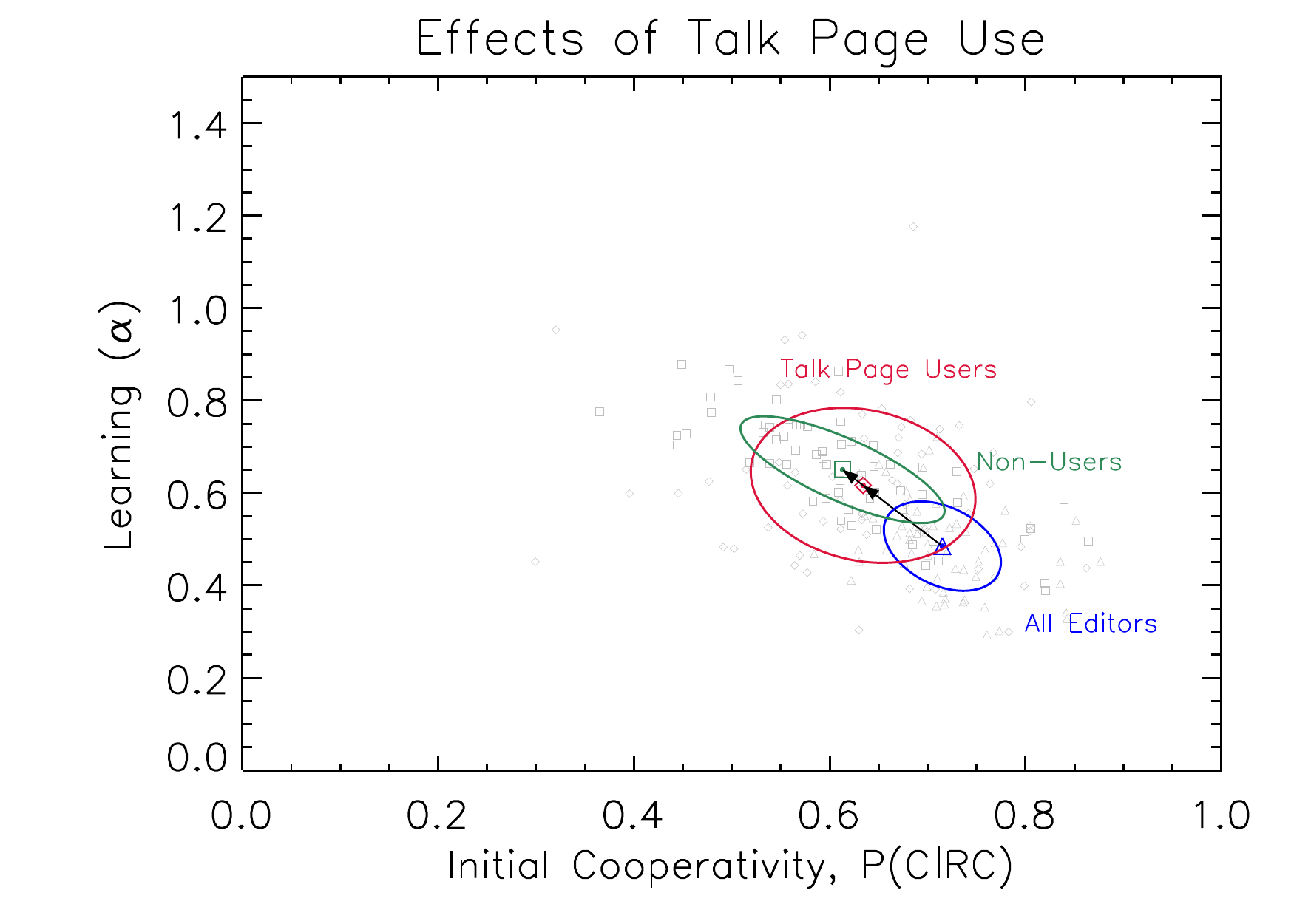}
\caption{{\bf Talking does not predict cooperation}. {\rm Users who have edited the article talk page recently (within the last ten edits on the mainpage) are quite similar to the population as a whole. Shown: distribution of $p$ and $\alpha$ parameters for 62 pages in our data, for all editors (triangles; blue $1\sigma$ range), recent talk page users (diamonds; red $1\sigma$ range) and non-users (squares; green $1\sigma$ range).}}
\label{int_talk}
\end{figure}
We also consider the effect of ``Talk Pages''---associated pages explicitly devoted to the discussion of contentious issues in the page itself. Fig.~\ref{int_talk} shows, for 62 pages in our data set, how $\alpha$ and $P(C|RC)$ shift when one considers the population as a whole, and two sub-populations: of editors who have recently (within the last ten edits) made an edit to the corresponding talk page, and those who have not.

Surprisingly, and in contrast to the findings for on-page interaction, we see little difference in behavior between the group of those who have, and those who have not, recently used the talk page. A similar result applies when we extend this window further to include talk page interaction within the past one-hundred edits. Talk does not shift behavior: though we emphasize, in keeping with our focus on group-level dynamics, that this conclusion applies to the joint state of the ``talk page users'' population, and not to any particular user.

Indeed, it is likely that individual users shift their opinions and beliefs a great deal through the unstructured discussion possible on talk pages. However, these shifts appear not to add together in a coherent fashion and do not leave a strong trace at higher, and more coarse-grained, levels of organization. What shifts preferences and beliefs at the group level is not talk, it seems, but action.

\section{Conclusion}
\label{conc}

This paper has worked through a minimal model of interaction on a large-scale networked social system. We showed how game-theoretic notions provided us a rigorous framework within which to move from observations of behavior to knowledge about beliefs and other mental states. The complex and networked structure of the system strongly suggests that we attribute these states to group-level cognitive processes.

Our basic model for interaction locates the struggle for sociality on Wikipedia in the extent to which users perceive the problem as a one-shot prisoner's dilemma, as opposed to a mutualistic interaction where the greatest benefits accrue to joint norm-following. This places our account of Wikipedia on the continuum between those who identify the origin of social complexity in altruism and group selection, vs.~those who see the burden on the side of cognitive ability (see, \emph{e.g.}, extended discussion in \citet{tomasello2009we,bowles2011cooperative}).

The recursive nature of both problems means that it not only the perception of users that matters, but their beliefs about the perceptions of others, and so forth. The fixed point solution we describe in Sec.~\ref{asg} implicitly assumes the existence of common knowledge. Real-world studies show that, at best, individuals only approximate such solutions \cite{frey2013cyclic}; indeed, results of this form, as well as theoretical arguments against the existence of true common knowledge \cite{gintis2014bounds}, strengthen our thesis two above---not only will individuals have more complex beliefs and shifting desires than the group-level beliefs, but they will reason in some bounded-rational fashion when deciding to act.

That group-level cognitive states appear to obey reasonably simple laws means not only that outside, scientific observers can form parsimonious descriptions of the system. It also means that such knowledge is potentially available to individuals within the system. Reference to ``how cooperative the group is'' will  likely form part of an individual user's reasoning. Such forms of group-level-pattern recognition are known to play a role in both human \cite{gigerenzer1999simple} and animal sociality \cite{Daniels28082012}. 

As shown in Sec.~\ref{beyond}, minds may be nested within minds. Discovery of such subgroup structure parallels the distinct subgroup strategies found in animal sociality \cite{dedeo2011evidence}, where triadic and higher-level decision-making is likely to be active \cite{dedeo2010inductive}.

Our results here are complementary to recent work on the quantification of group agency established in a series of influential papers by List and Pettit \cite{lp1, lp2, lp3}. This work establishes that while some limit cases may be trivial (everyone believes X, and so ``the group believes X''), theorems on preference aggregation mean that the relationship between individual and group-level mental states is complex. Under List and Pettit's analysis, some groups may fail to be true agents able to form intentional beliefs (``minds,'' in our language) because of the relations between the individuals that compose these groups. For List and Pettit, the ascription requires that individuals dynamically respond to each other in sufficiently cooperative ways. We have no such constraint, but anticipate that only in groups with sufficiently aligned incentives, or sufficient commitments to norm creation and rule-following, will large-scale cognitive laws be useful sources of knowledge.

Our account is complementary to an influential strand of thought that locates the boundaries of mind---and, in particular, of unitary experience---by reference to the degree of integration between subparts \cite{Tononi24051994,Tononi1998474,oizumi2014phenomenology}. There is no question in this case that, at least on longer timescales (hours to days), users do not interact frequently enough via the Wikipedia interface to justify a claim that these cognitive states are accompanied by qualitative experience. The interaction between users and distributed, higher-level properties associated with group-level beliefs has important commonalities with current thinking on the emergence of life \cite{Walker06022013, walker2012evolutionary, hoel2013quantifying}, and with accounts of multi-level selection in evolutionary systems \cite{traulsen2006evolution}.

Our work connects directly to philosophical accounts of the Group Mind hypothesis \cite{Theiner2008-THEFEM,Theiner2010-THERGC,Theiner2013-THEOAU}, in as much as it takes the cognitive states of these larger, non-biological units seriously. When individuals interact in sufficiently complex ways, the same reasons for including features of the external, physical world as part of a cognitive process \cite{Clark01011998,clark2008supersizing} may apply to the inclusion of the cognitive states of others. These aggregate processes---group minds---will naturally have mental properties that differ from the individual minds out of which they are constructed.

The existence of group-level cognitive states argues for cognitive approaches to the evolution of societies, including the quantification of their inferential and information-theoretic properties \cite{dedeo2013bootstrap}, potentially over timescales longer than an individual lifespan \cite{klingenstein2014civilizing}. Our work is thus of direct relevance to a long-standing debate in the social and historical sciences \cite{elias2000society}. What is the natural unit of analysis for long-term historical change, or the behavior of large groups? Methodological individualism necessarily reduces all talk of group-level beliefs, including accounts of culture and institutions, to the individual \cite{bowles2009microeconomics,bowles2011cooperative}. From Sec.~\ref{jm} we know that it is (of course) not impossible to give an account of group-level mental states (the macrostate) in terms of individual mental states (the microstate)---only that, as we see in Sec.~\ref{complex}, it may become extremely unwieldy, requiring us to explain simple laws in terms of an enormous number of fine-tuned processes. 

\section{Acknowledgements}
\noindent
I thank the organizers and attendees of the ASU BEYOND Center February 2014 meeting ``Information, Complexity and Life'', of the 2014 Santa Fe Institute (SFI) Complex Systems Summer School, and of the 2014 SFI Graduate Workshop in Computational Social Science, for useful discussions, and my two reviewers, Theodore P. Pavlic (Arizona State University) and an anonymous referee, for detailed feedback on this manuscript. I thank the City University of New York's Initiative for the Theoretical Sciences for their hospitality while this work was completed. This work was supported in part by National Science Foundation Grant EF-1137929. 

\section{Appendix: Measuring $\alpha$ and $p$}

In \citet{dedeo2013collective}, the relevant parameters $p$ and $\alpha$, were estimated from the counts of strings of the form $RC^kR$, under the assumption that events of the form $RC^kR$ are Poisson-distributed with mean varying as a function of $k$. Here, we present a simpler method for the estimation of $\alpha$ and $p$ from data. In particular, we estimate $p$ and $\alpha$ from observations of $N(R|RC^k)$; using the binomial distribution, we maximize the posterior
\begin{eqnarray}
\log{P(\textrm{data}|p,\alpha)} & = & \sum_{i=1}^\infty \left(\log{n_i!}-\log{k_i!}-\log{(n_i-k_i)!}\right) \nonumber \\
& & {}+ k_i\log{\left(\frac{p}{(i+1)^\alpha}\right)} + (n_i-k_i)\log{\left(1-\frac{p}{(i+1)^\alpha}\right)},
\label{measuring}
\end{eqnarray}
where $n_i$ is the number of times one observes a string of length $RC^i$, and $k_i$ is the number of times one observes a string of length $RC^iR$. In words, each term in the sum is the log-likelihood of a string $RC^i$ being followed by a revert, given the choice of $p$ and $\alpha$.

Eq.~\ref{measuring} gives slightly different values for $p$ and $\alpha$ than the estimation method of \citet{dedeo2013collective}. In addition to being easier to compute, it makes fewer assumptions about the process---in particular, it does not require one assume that the number of sequences of the form $RC^kR$ is much less than the total length of the series.

Using Eq.~\ref{measuring}, it is simple to estimate more complicated conditional probabilities, as required in Sec.~\ref{beyond}. Consider the probability that an editor performs a revert action, given (1) $k$ cooperative events have been seen and (2) the editor has previously interacted with the page,
\begin{equation}
P(R|RC^k,\mathrm{int}) = \frac{p_\mathrm{int}}{(k+1)^{\alpha_\mathrm{int}}}.
\end{equation}
One can find the maximum a posteriori values of $p_\mathrm{int}$ and $\alpha_\mathrm{int}$ by reference to $n_{i,\mathrm{int}}$ and $k_{i,\mathrm{int}}$, where $n_{i,\mathrm{int}}$ is the number of times a cooperative sequence of length $i$ is followed by a $C$ or $R$ move by an editor who has previously interacted with the page, and $k_{i,\mathrm{int}}$ is the number of times a cooperative sequence of length $i$ is terminated by an $R$ move by an editor who has previously interacted with the page.

\section{Appendix: The Distant Revert Problem}

Detecting reverts by looking for the exact replication of page content (``hash-reverts'') across two edits is a standard method in research into Wiki-like interactions. However, such a technique may miss an important feature of ``undo-like'' conflict.\footnote{I am grateful to one of my referees, Theodore P. Pavlic, for drawing this feature of Wiki-like conflict to my attention.} If we index all ``C'' edits by the sum total of content changes they make, then we can index ``R''-type edits by the set of ``C'' edits they undo. For example, consider the following sequence of page contents:
\begin{verse}
Version 1, User A \\
{\tt George Washington was our first president.}
\end{verse}
\begin{verse}
Version 2, User B \\
{\tt George Washington was our first president.} \\
{\tt Famously, he cut down a cherry tree while young.}
\end{verse}
\begin{verse}
Version 3, User C \\
{\tt George Washington was our first president.} \\
{\tt Famously, he cut down a cherry tree while young.} \\
{\tt He warned against the formation of political parties.}
\end{verse}
\begin{verse}
Version 4, User D \\
{\tt George Washington was our first president.} \\
{\tt He warned against the formation of political parties.}
\end{verse}
In the standard method of tracking revert behavior, this sequence contains no hash-reverts---no complete page text appears twice---and would be recorded as $CCCC$. However, notice that User D, while not undoing the work of User C, does undo the work of User B: she deletes the apocryphal story of Washington cutting down the cherry tree. It is reasonable to imagine tracking this sequence as $C_1C_2C_3R_2$--\emph{i.e.}, the last edit as a reversion of the edit $C_2$. Depending on the relative rates at which users make edits, and other users disagree with those edits and undo them, the dynamics of distant reverts can obey very different laws from those of hash-reverts; indeed, it is possible for distant reverts to be uncorrelated (\emph{i.e.}, follow an exponential distribution with $\alpha$ equal to zero in Eq.~\ref{phenom}) while hash-reverts appear correlated.

Tracking of these distant reverts is non-trivial; it is most easily done by defining an edit as a set of minimum-Levenshtein distance edits, using a tool such as UNIX {\tt diff}. We can then define a distant revert as a new edit whose diff is the exact opposite of a previous edit: all additions become deletions, all deletions become additions, and all changes are reversed. An example of a distant revert is the pair of edits made by users {\tt Wiki alf} and {\tt Joker123192} to the {\tt George\_W.\_Bush} page; on 18 May 2009, {\tt Wiki alf} adds a paragraph; and on 29 May 2009, 32 edits later, {\tt Joker123192} takes it back out.\footnote{See \url{https://en.wikipedia.org/w/index.php?title=Main_Page&diff=290774306&oldid=290770322} and \url{https://en.wikipedia.org/w/index.php?title=Main_Page&diff=293157787&oldid=292576084}, last accessed 8 September 2014.}

For the {\tt George\_W.\_Bush} page, inclusion of these single-index reverts increases the number of reverts counted from 14,976 to 15,982; this shifts $\alpha$ from $0.41\pm0.02$ to $0.38\pm0.02$. UNIX {\tt diff} works newline-by-newline, so that if Edit 1 adds a sentence, Edit 2 adds a sentence to the same paragraph, and Edit 3 deletes the sentence added in Edit 1, this will appear as a change, rather than a deletion. One can instead work sentence-by-sentence, to look for potential disagreements about content internal to a paragraph. Doing so finds a total of 16,021 reverts, and (within error) does not change the estimate of $\alpha$. Note our estimate of $\alpha$ for the {\tt George\_W.\_Bush} differs from that in~\citet{dedeo2013collective}; this is both because we include more data (through 4 July 2014), and because we use an improved method for the estimation of $p$ and $\alpha$.

Strong norms exist against hash-reverts. For example, the idea that a revert leads to the repetition of prior page content is often part of the definition of revert on Wikipedia itself.\footnote{``Reverting means undoing or otherwise negating the effects of one or more edits, which results in the page being restored to a previous version.'' (see \url{http://en.wikipedia.org/wiki/Help:Reverting}). A weaker claim appears at \url{http://en.wikipedia.org/wiki/Wikipedia:Reverting}: ``Reverting means reversing a prior edit, which typically results in the article being restored to a version that existed sometime previously.'' Both accessed 8 September 2014.} By contrast, Wikipedia norms are more ambiguous when it comes to distant reverts, and these reverts may be weaker signals of conflict. For intuition as to why, consider the following sequence:
\begin{verse}
Version 1, User A \\
{\tt George Washington was our first president.}
\end{verse}
\begin{verse}
Version 2, User B \\
{\tt George Washington was our first president.} \\
{\tt He was against political polarization.}
\end{verse}
\begin{verse}
Version 3, User C \\
{\tt George Washington was our first president.} \\
{\tt He was against political polarization.} \\
{\tt He warned against the formation of political parties.}
\end{verse}
\begin{verse}
Version 4, User D \\
{\tt George Washington was our first president.} \\
{\tt He warned against the formation of political parties.}
\end{verse}
Here, User D has partially reverted User B, as before; but (it may be inferred) in part because she might believe that User's C's addition renders User B's contribution moot, rather than because she believes User B's content to be unambiguously bad. If User B disagrees with User D, he may roll back to Version 3 or to Version 2; in either case, this will now appear as a hash-revert. 


\section{Appendix: Wikipedia pages used in this analysis}

2006\_Lebanon\_War, Argentina, Atheism, Australia, Barack\_Obama, Blackout\_(Britney\_Spears\-\_album), Blink-182, Bob\_Dylan, Canada, Catholic\_Church, Che\_Guevara, China, Circumcision, Cuba, Eminem, France, Gaza\_War, Genghis\_Khan, George\_W.\_Bush, Girls'\_Generation, Global\-\_warming, God, Golf, Heroes\_(TV\_series), Hilary\_Duff, Hillary\_Rodham\_Clinton, Homosexuality, Hurricane\_Katrina, IPhone, Iraq\_War, Islam, Israel, John\_F.\_Kennedy, John\_Kerry, Lindsay\-\_Lohan, Linux, Lost\_(TV\_series), Mexico, Michael\_Jackson, Neighbours, New\_Zealand, Paris\-\_Hilton, Paul\_McCartney, Pink\_Floyd, RMS\_Titanic, Russo-Georgian\_War, Scientology, September\_11\_attacks, Shakira, Star\_Trek, Super\_Smash\_Bros.\_Brawl, Sweden, The\_Dark\_Knight\_\-(film), The\_Holocaust, Turkey, United\_States, Virginia\_Tech\_massacre, Wiki\-pedia, Wizards\_of\-\_Waverly\_\-Place, World\_War\_I, World\_War\_II, Xbox\_360

\bibliography{new_joint}


\begin{thebibliography}{00}


\ifx \showCODEN    \undefined \def \showCODEN     #1{\unskip}     \fi
\ifx \showDOI      \undefined \def \showDOI       #1{{\tt DOI:}\penalty0{#1}\ }
  \fi
\ifx \showISBNx    \undefined \def \showISBNx     #1{\unskip}     \fi
\ifx \showISBNxiii \undefined \def \showISBNxiii  #1{\unskip}     \fi
\ifx \showISSN     \undefined \def \showISSN      #1{\unskip}     \fi
\ifx \showLCCN     \undefined \def \showLCCN      #1{\unskip}     \fi
\ifx \shownote     \undefined \def \shownote      #1{#1}          \fi
\ifx \showarticletitle \undefined \def \showarticletitle #1{#1}   \fi
\ifx \showURL      \undefined \def \showURL       #1{#1}          \fi

\bibitem[\protect\citeauthoryear{Arrow}{Arrow}{1994}]%
        {arrow1994}
{Arrow, K.~J}. (1994).
\newblock \showarticletitle{Methodological Individualism and Social Knowledge}.
\newblock {\em The American Economic Review\/} {84}, 2 (1994), pp. 1--9.
\newblock
\showISSN{00028282}


\bibitem[\protect\citeauthoryear{Bar-Ilan and Aharony}{Bar-Ilan and
  Aharony}{2014}]%
        {bar2014twelve}
{Bar-Ilan, J} {and} {Aharony, N}. (2014).
\newblock \showarticletitle{Twelve years of {W}ikipedia research}. In {\em
  Proceedings of the 2014 ACM conference on Web science}. 243--244.
\newblock


\bibitem[\protect\citeauthoryear{Bowles}{Bowles}{2009}]%
        {bowles2009microeconomics}
{Bowles, S}. (2009).
\newblock {\em {Microeconomics: Behavior, Institutions, and Evolution}}.
\newblock Princeton University Press.
\newblock


\bibitem[\protect\citeauthoryear{Bowles and Gintis}{Bowles and Gintis}{2011}]%
        {bowles2011cooperative}
{Bowles, S} {and} {Gintis, H}. (2011).
\newblock {\em {A Cooperative Species: Human Reciprocity and its Evolution}}.
\newblock Princeton University Press.
\newblock


\bibitem[\protect\citeauthoryear{Clark}{Clark}{2008}]%
        {clark2008supersizing}
{Clark, A}. (2008).
\newblock {\em Supersizing the Mind: Embodiment, Action, and Cognitive
  Extension}.
\newblock Oxford University Press.
\newblock


\bibitem[\protect\citeauthoryear{Clark and Chalmers}{Clark and
  Chalmers}{1998}]%
        {Clark01011998}
{Clark, A} {and} {Chalmers, D}. (1998).
\newblock \showarticletitle{The Extended Mind}.
\newblock {\em Analysis\/} {58}, 1 (1998), 7--19.
\newblock
\showDOI{%
\url{http://dx.doi.org/10.1093/analys/58.1.7}}


\bibitem[\protect\citeauthoryear{Crutchfield and Young}{Crutchfield and
  Young}{1989}]%
        {crutchfield1989inferring}
{Crutchfield, J.~P} {and} {Young, K}. (1989).
\newblock \showarticletitle{Inferring statistical complexity}.
\newblock {\em Physical Review Letters\/} {63}, 2 (1989), 105.
\newblock


\bibitem[\protect\citeauthoryear{Daniels, Krakauer, and Flack}{Daniels
  et~al\mbox{.}}{2012}]%
        {Daniels28082012}
{Daniels, B.~C}, {Krakauer, D.~C}, {and} {Flack, J.~C}. (2012).
\newblock \showarticletitle{Sparse code of conflict in a primate society}.
\newblock {\em Proceedings of the National Academy of Sciences\/} {109}, 35
  (2012), 14259--14264.
\newblock
\showDOI{%
\url{http://dx.doi.org/10.1073/pnas.1203021109}}


\bibitem[\protect\citeauthoryear{DeDeo}{DeDeo}{2011}]%
        {dedeo2011effective}
{DeDeo, S}. (2011).
\newblock \showarticletitle{Effective theories for circuits and automata}.
\newblock {\em Chaos\/} {21}, 3 (2011), 037106.
\newblock
\showDOI{%
\url{http://dx.doi.org/10.1063/1.3640747}}


\bibitem[\protect\citeauthoryear{DeDeo}{DeDeo}{2013}]%
        {dedeo2013collective}
{DeDeo, S}. (2013).
\newblock \showarticletitle{Collective Phenomena and Non-Finite State
  Computation in a Human Social System}.
\newblock {\em PLoS one\/} {8}, 10 (2013), e75818.
\newblock


\bibitem[\protect\citeauthoryear{DeDeo, Hawkins, Klingenstein, and
  Hitchcock}{DeDeo et~al\mbox{.}}{2013}]%
        {dedeo2013bootstrap}
{DeDeo, S}, {Hawkins, R.~X}, {Klingenstein, S}, {and} {Hitchcock, T}. (2013).
\newblock \showarticletitle{Bootstrap methods for the empirical study of
  decision-making and information flows in social systems}.
\newblock {\em Entropy\/} {15}, 6 (2013), 2246--2276.
\newblock


\bibitem[\protect\citeauthoryear{DeDeo, Krakauer, and Flack}{DeDeo
  et~al\mbox{.}}{2010}]%
        {dedeo2010inductive}
{DeDeo, S}, {Krakauer, D}, {and} {Flack, J}. (2010).
\newblock \showarticletitle{Inductive game theory and the dynamics of animal
  conflict}.
\newblock {\em PLoS computational biology\/} {6}, 5 (2010), e1000782.
\newblock


\bibitem[\protect\citeauthoryear{DeDeo, Krakauer, and Flack}{DeDeo
  et~al\mbox{.}}{2011}]%
        {dedeo2011evidence}
{DeDeo, S}, {Krakauer, D}, {and} {Flack, J}. (2011).
\newblock \showarticletitle{Evidence of strategic periodicities in collective
  conflict dynamics}.
\newblock {\em Journal of The Royal Society Interface\/} {8}, 62 (2011),
  1260--1273.
\newblock


\bibitem[\protect\citeauthoryear{Elias}{Elias}{2000a}]%
        {elias2000civilizing}
{Elias, N}. (2000)a.
\newblock {\em The Civilizing Process: Sociogenetic and Psychogenetic
  Investigations}.
\newblock Wiley.
\newblock
\showISBNx{9780631221616}
\showLCCN{99056203}
\newblock
\shownote{Second Edition of 1939 Text, edited by Dunning, E., Goudsblom, J.,
  and Mennell, S.}


\bibitem[\protect\citeauthoryear{Elias}{Elias}{2000b}]%
        {elias2000society}
{Elias, N}. (2000)b.
\newblock {\em The Society of Individuals}.
\newblock Wiley.
\newblock
\showISBNx{9780631221616}
\showLCCN{99056203}
\newblock
\shownote{Edited by Michael Schroter, translated by Edmund Jephcott; first
  edition, 1987.}


\bibitem[\protect\citeauthoryear{Frey and Goldstone}{Frey and
  Goldstone}{2013}]%
        {frey2013cyclic}
{Frey, S} {and} {Goldstone, R.~L}. (2013).
\newblock \showarticletitle{Cyclic game dynamics driven by iterated reasoning}.
\newblock {\em PLoS one\/} {8}, 2 (2013), e56416.
\newblock


\bibitem[\protect\citeauthoryear{Fudenberg and Tirole}{Fudenberg and
  Tirole}{1991}]%
        {fudenberg1991game}
{Fudenberg, D} {and} {Tirole, J}. (1991).
\newblock {\em Game Theory}.
\newblock MIT Press.
\newblock
\showISBNx{9780262061414}
\showLCCN{91002301}


\bibitem[\protect\citeauthoryear{Gigerenzer and Todd}{Gigerenzer and
  Todd}{1999}]%
        {gigerenzer1999simple}
{Gigerenzer, G} {and} {Todd, P.~M}. (1999).
\newblock {\em Simple heuristics that make us smart}.
\newblock Oxford University Press.
\newblock


\bibitem[\protect\citeauthoryear{Gintis}{Gintis}{2014}]%
        {gintis2014bounds}
{Gintis, H}. (2014).
\newblock {\em The Bounds of Reason: Game Theory and the Unification of the
  Behavioral Sciences}.
\newblock Princeton University Press.
\newblock
\showISBNx{9781400851348}
\newblock
\shownote{Second edition.}


\bibitem[\protect\citeauthoryear{Greenstein and Zhu}{Greenstein and
  Zhu}{2012}]%
        {greenstein2012wikipedia}
{Greenstein, S} {and} {Zhu, F}. (2012).
\newblock \showarticletitle{Is {W}ikipedia Biased?}
\newblock {\em The American Economic Review\/} {102}, 3 (2012), 343--348.
\newblock


\bibitem[\protect\citeauthoryear{Halfaker, Geiger, Morgan, and Riedl}{Halfaker
  et~al\mbox{.}}{2012}]%
        {halfaker2012rise}
{Halfaker, A}, {Geiger, R.~S}, {Morgan, J.~T}, {and} {Riedl, J}. (2012).
\newblock \showarticletitle{{The rise and decline of an open collaboration
  system: How {W}ikipedia's reaction to popularity is causing its decline}}.
\newblock {\em American Behavioral Scientist\/} {57}, 5 (2012), 665--688.
\newblock
\showDOI{%
\url{http://dx.doi.org/10.1177/0002764212469365}}


\bibitem[\protect\citeauthoryear{Hayek}{Hayek}{1945}]%
        {hayek1945use}
{Hayek, F.~A}. (1945).
\newblock \showarticletitle{The use of knowledge in society}.
\newblock {\em The American Economic Review\/} {35}, 4 (1945), 519--530.
\newblock


\bibitem[\protect\citeauthoryear{Hazard}{Hazard}{2013}]%
        {hazard2013crisis}
{Hazard, P}. (2013).
\newblock {\em {The crisis of the European mind, 1680--1715}}.
\newblock New York Review of Books.
\newblock


\bibitem[\protect\citeauthoryear{Hoel, Albantakis, and Tononi}{Hoel
  et~al\mbox{.}}{2013}]%
        {hoel2013quantifying}
{Hoel, E.~P}, {Albantakis, L}, {and} {Tononi, G}. (2013).
\newblock \showarticletitle{Quantifying causal emergence shows that macro can
  beat micro}.
\newblock {\em Proceedings of the National Academy of Sciences\/} {110}, 49
  (2013), 19790--19795.
\newblock


\bibitem[\protect\citeauthoryear{Jemielniak}{Jemielniak}{2014}]%
        {jemielniak2014common}
{Jemielniak, D}. (2014).
\newblock {\em Common Knowledge?: An Ethnography of {W}ikipedia}.
\newblock Stanford University Press.
\newblock
\showISBNx{9780804789448}
\showLCCN{2013047786}


\bibitem[\protect\citeauthoryear{Kiesler, Kraut, Resnick, and Kittur}{Kiesler
  et~al\mbox{.}}{2012}]%
        {regulating}
{Kiesler, S}, {Kraut, R}, {Resnick, P}, {and} {Kittur, A}. (2012).
\newblock \showarticletitle{Regulating Behavior in Online Communities}.
\newblock In {\em Building Successful Online Communities: Evidence-Based Social
  Design}, {R.E. Kraut}, {P.~Resnick}, {S.~Kiesler}, {M.~Burke}, {Y.~Chen},
  {N.~Kittur}, {J.~Konstan}, {Y.~Ren}, {and} {J.~Riedl} (Eds.). MIT Press.
\newblock
\showISBNx{9780262297394}


\bibitem[\protect\citeauthoryear{Kirman}{Kirman}{2010}]%
        {kirman2010complex}
{Kirman, A}. (2010).
\newblock {\em Complex economics: individual and collective rationality}.
\newblock Routledge.
\newblock


\bibitem[\protect\citeauthoryear{Kittur and Kraut}{Kittur and Kraut}{2010}]%
        {kittur2010beyond}
{Kittur, A} {and} {Kraut, R.~E}. (2010).
\newblock \showarticletitle{Beyond Wikipedia: coordination and conflict in
  online production groups}. In {\em {Proceedings of the 2010 ACM conference on
  Computer Supported Cooperative Work}}. ACM, 215--224.
\newblock


\bibitem[\protect\citeauthoryear{Klingenstein, Hitchcock, and
  DeDeo}{Klingenstein et~al\mbox{.}}{2014}]%
        {klingenstein2014civilizing}
{Klingenstein, S}, {Hitchcock, T}, {and} {DeDeo, S}. (2014).
\newblock \showarticletitle{{The civilizing process in London's Old Bailey}}.
\newblock {\em Proceedings of the National Academy of Sciences\/} {111}, 26
  (2014), 9419--9424.
\newblock


\bibitem[\protect\citeauthoryear{Ko{\l}akowski}{Ko{\l}akowski}{2008}]%
        {marxcurrents}
{Ko{\l}akowski, L}. (2008).
\newblock {\em Main Currents of Marxism}.
\newblock W.~W.~Norton \& Company.
\newblock
\showISBNx{978-0393329438}
\newblock
\shownote{Translated by P.~S.~Falla; first edition, 1976.}


\bibitem[\protect\citeauthoryear{Kreps}{Kreps}{1990}]%
        {kreps1990game}
{Kreps, D}. (1990).
\newblock {\em Game Theory and Economic Modelling}.
\newblock Clarendon Press, Oxford, UK.
\newblock
\showISBNx{9780198283812}
\showLCCN{91118944}


\bibitem[\protect\citeauthoryear{Lam and Riedl}{Lam and Riedl}{2009}]%
        {tails}
{Lam, S.~T.~K} {and} {Riedl, J}. (2009).
\newblock \showarticletitle{Is {W}ikipedia Growing a Longer Tail?}. In {\em
  Proceedings of the ACM 2009 International Conference on Supporting Group
  Work} {\em (GROUP '09)}. ACM, New York, NY, USA, 105--114.
\newblock
\showISBNx{978-1-60558-500-0}
\showDOI{%
\url{http://dx.doi.org/10.1145/1531674.1531690}}


\bibitem[\protect\citeauthoryear{List and Pettit}{List and Pettit}{2002}]%
        {lp1}
{List, C} {and} {Pettit, P}. (2002).
\newblock \showarticletitle{Aggregating sets of judgments: An impossibility
  result}.
\newblock {\em Economics and Philosophy\/} {18}, 01 (2002), 89--110.
\newblock


\bibitem[\protect\citeauthoryear{List and Pettit}{List and Pettit}{2004}]%
        {lp2}
{List, C} {and} {Pettit, P}. (2004).
\newblock \showarticletitle{Aggregating Sets of Judgments: Two Impossibility
  Results Compared}.
\newblock {\em Synthese\/} {140}, 1-2 (2004), 207--235.
\newblock


\bibitem[\protect\citeauthoryear{List and Pettit}{List and Pettit}{2011}]%
        {lp3}
{List, C} {and} {Pettit, P}. (2011).
\newblock {\em Group agency: The possibility, design, and status of corporate
  agents}.
\newblock Oxford University Press.
\newblock


\bibitem[\protect\citeauthoryear{Minsky}{Minsky}{1967}]%
        {minsky1976computation}
{Minsky, M}. (1967).
\newblock {\em Computation: Finite and Infinite Machines}.
\newblock Prentice-Hall.
\newblock


\bibitem[\protect\citeauthoryear{Oizumi, Albantakis, and Tononi}{Oizumi
  et~al\mbox{.}}{2014}]%
        {oizumi2014phenomenology}
{Oizumi, M}, {Albantakis, L}, {and} {Tononi, G}. (2014).
\newblock \showarticletitle{From the Phenomenology to the Mechanisms of
  Consciousness: Integrated Information Theory 3.0}.
\newblock {\em PLoS computational biology\/} {10}, 5 (2014), e1003588.
\newblock


\bibitem[\protect\citeauthoryear{Reagle}{Reagle}{2010}]%
        {reagle2010good}
{Reagle, J.~M}. (2010).
\newblock {\em Good Faith Collaboration: The Culture of {W}ikipedia}.
\newblock MIT Press.
\newblock
\showISBNx{9780262014472}
\showLCCN{2009052779}


\bibitem[\protect\citeauthoryear{San\'{i}n and Wood}{San\'{i}n and
  Wood}{2014}]%
        {Sanin01032014}
{San\'{i}n, F.~G} {and} {Wood, E.~J}. (2014).
\newblock \showarticletitle{Ideology in civil war: Instrumental adoption and
  beyond}.
\newblock {\em Journal of Peace Research\/} {51}, 2 (2014), 213--226.
\newblock
\showDOI{%
\url{http://dx.doi.org/10.1177/0022343313514073}}


\bibitem[\protect\citeauthoryear{Shalizi and Moore}{Shalizi and Moore}{2003}]%
        {shalizi2003macrostate}
{Shalizi, C.~R} {and} {Moore, C}. (2003).
\newblock \showarticletitle{What is a macrostate? Subjective observations and
  objective dynamics}.
\newblock {\em arXiv preprint cond-mat/0303625\/} (2003).
\newblock


\bibitem[\protect\citeauthoryear{Steeg, Galstyan, Sha, and DeDeo}{Steeg
  et~al\mbox{.}}{2014}]%
        {steeg2013demystifying}
{Steeg, G.~V}, {Galstyan, A}, {Sha, F}, {and} {DeDeo, S}. (2014).
\newblock \showarticletitle{Demystifying Information-Theoretic Clustering}.
\newblock {\em Journal of Machine Learning Research\/} {31}, 1 (2014), 19--27.
\newblock
\newblock
\shownote{Proceedings of The 31st International Conference on Machine Learning;
  arXiv:1310.4210.}


\bibitem[\protect\citeauthoryear{Theiner}{Theiner}{2008}]%
        {Theiner2008-THEFEM}
{Theiner, G}. (2008).
\newblock {\em From Extended Minds to Group Minds: Rethinking the Boundaries of
  the Mental}.
\newblock Ph.D. Dissertation. Indiana University.
\newblock


\bibitem[\protect\citeauthoryear{Theiner}{Theiner}{2013}]%
        {Theiner2013-THEOAU}
{Theiner, G}. (2013).
\newblock \showarticletitle{Onwards and Upwards with the Extended Mind: From
  Individual to Collective Epistemic Action}.
\newblock In {\em Developing Scaffolds}, {Linnda Caporael}, {James Griesemer},
  {and} {William Wimsatt} (Eds.). MIT Press, 191--208.
\newblock


\bibitem[\protect\citeauthoryear{Theiner, Allen, and Goldstone}{Theiner
  et~al\mbox{.}}{2010}]%
        {Theiner2010-THERGC}
{Theiner, G}, {Allen, C}, {and} {Goldstone, R.~L}. (2010).
\newblock \showarticletitle{Recognizing Group Cognition}.
\newblock {\em Cognitive Systems Research\/} {11}, 4 (2010), 378--395.
\newblock


\bibitem[\protect\citeauthoryear{Tomasello}{Tomasello}{2009}]%
        {tomasello2009we}
{Tomasello, M}. (2009).
\newblock {\em Why We Cooperate}.
\newblock MIT Press.
\newblock
\showISBNx{9780262258494}
\newblock
\shownote{Responses by Carol Dweck, Joan Silk, Brian Skyrms, and Elizabeth
  Spelke.}


\bibitem[\protect\citeauthoryear{Tononi, Edelman, and Sporns}{Tononi
  et~al\mbox{.}}{1998}]%
        {Tononi1998474}
{Tononi, G}, {Edelman, G.~M}, {and} {Sporns, O}. (1998).
\newblock \showarticletitle{Complexity and coherency: integrating information
  in the brain}.
\newblock {\em Trends in Cognitive Sciences\/} {2}, 12 (1998), 474 -- 484.
\newblock
\showISSN{1364-6613}
\showDOI{%
\url{http://dx.doi.org/10.1016/S1364-6613(98)01259-5}}


\bibitem[\protect\citeauthoryear{Tononi, Sporns, and Edelman}{Tononi
  et~al\mbox{.}}{1994}]%
        {Tononi24051994}
{Tononi, G}, {Sporns, O}, {and} {Edelman, G}. (1994).
\newblock \showarticletitle{A measure for brain complexity: relating functional
  segregation and integration in the nervous system}.
\newblock {\em Proceedings of the National Academy of Sciences\/} {91}, 11
  (1994), 5033--5037.
\newblock
\showURL{%
\url{http://www.pnas.org/content/91/11/5033.abstract}}


\bibitem[\protect\citeauthoryear{Towne, Kittur, Kinnaird, and Herbsleb}{Towne
  et~al\mbox{.}}{2013}]%
        {kittur}
{Towne, W.~B}, {Kittur, A}, {Kinnaird, P}, {and} {Herbsleb, J.~D}. (2013).
\newblock \showarticletitle{Your process is showing: controversy management and
  perceived quality in {W}ikipedia}. In {\em Computer Supported Cooperative
  Work, CSCW 2013, San Antonio, TX, USA, February 23-27, 2013}, {Amy Bruckman},
  {Scott Counts}, {Cliff Lampe}, {and} {Loren~G. Terveen} (Eds.). ACM,
  1059--1068.
\newblock
\showISBNx{978-1-4503-1331-5}


\bibitem[\protect\citeauthoryear{Traulsen and Nowak}{Traulsen and
  Nowak}{2006}]%
        {traulsen2006evolution}
{Traulsen, A} {and} {Nowak, M.~A}. (2006).
\newblock \showarticletitle{Evolution of cooperation by multilevel selection}.
\newblock {\em Proceedings of the National Academy of Sciences\/} {103}, 29
  (2006), 10952--10955.
\newblock


\bibitem[\protect\citeauthoryear{Walker, Cisneros, and Davies}{Walker
  et~al\mbox{.}}{2012}]%
        {walker2012evolutionary}
{Walker, S.~I}, {Cisneros, L}, {and} {Davies, P.~C}. (2012).
\newblock \showarticletitle{Evolutionary transitions and top-down causation}.
\newblock {\em arXiv preprint arXiv:1207.4808\/} (2012).
\newblock


\bibitem[\protect\citeauthoryear{Walker and Davies}{Walker and Davies}{2013}]%
        {Walker06022013}
{Walker, S.~I} {and} {Davies, P.~C.~W}. (2013).
\newblock \showarticletitle{The algorithmic origins of life}.
\newblock {\em Journal of The Royal Society Interface\/} {10}, 79 (2013).
\newblock
\showDOI{%
\url{http://dx.doi.org/10.1098/rsif.2012.0869}}


\bibitem[\protect\citeauthoryear{Wegner}{Wegner}{1987}]%
        {wegner1987transactive}
{Wegner, D.~M}. (1987).
\newblock \showarticletitle{Transactive memory: A contemporary analysis of the
  group mind}.
\newblock In {\em Theories of group behavior}. Springer, 185--208.
\newblock


\bibitem[\protect\citeauthoryear{Welser, Cosley, Kossinets, Lin, Dokshin, Gay,
  and Smith}{Welser et~al\mbox{.}}{2011}]%
        {roles}
{Welser, H.~T}, {Cosley, D}, {Kossinets, G}, {Lin, A}, {Dokshin, F}, {Gay, G},
  {and} {Smith, M}. (2011).
\newblock \showarticletitle{Finding Social Roles in {W}ikipedia}. In {\em
  Proceedings of the 2011 iConference} {\em (iConference '11)}. ACM, New York,
  NY, USA, 122--129.
\newblock
\showISBNx{978-1-4503-0121-3}
\showDOI{%
\url{http://dx.doi.org/10.1145/1940761.1940778}}


\end{thebibliography}

\end{document}